\documentclass[twocolumn,showpacs,pra,raggedbottom]{revtex4}
\usepackage{graphicx}
\begin{document}

\title{Properties of quasi-one-dimensional molecules with Feshbach
resonance interaction}

\author{V. A. Yurovsky}

\affiliation{School of Chemistry, Tel Aviv University, 69978 Tel Aviv,
Israel}

\date{\today}

\begin{abstract}
Bound states and collisions of atoms with
two-channel two-body interactions in harmonic waveguides are
analyzed. The closed-channel contributions to two-atom bound
states become dominant in the case of a weak resonance. At low
energies and values of the non-resonant scattering length the
problem can be approximated by a one-dimensional resonant model.
Three-body problem becomes nonintegrable and the properties of
triatomic molecules become different from those predicted by the
integrable Lieb-Liniger-McGuire model.
\end{abstract}

\pacs{03.65.Ge, 03.65.Nk, 03.75.Be, 32.80.Pj}
\maketitle    

\section*{Introduction}

Quasi-one-dimensional (1D) molecules, the existence of which was
predicted in Ref.\ \cite{BMO03}, have been recently observed in an
 experiment
\cite{MSGKE05} with two-dimensional optical lattices. Except of
 lattices (see
also Refs.\ \cite{G01,W04,NIST_Latt05}), tight cylindrical
 confinements, or
atomic waveguides, have been realized in elongated atomic traps (see
 Refs.\
\cite{G01a,S02,K02,R03} and references therein), and atomic
 integrated optics
devices (see Refs.\ \cite{F02,MIT_Chip05} and references therein).
 Unlike
weakly bound molecules in free, three-dimensional (3D), space (see
 Refs.\
\cite{GKGTJ04,Chin05} and references therein), existing at positive
 elastic
scattering length only, the quasi-1D ones survive at negative
 scattering
length.

The interaction of two atoms in an atomic waveguide can be
considered as a 1D zero-range interaction (see Ref.\ \cite{O98})
whenever the collision or binding energies remain small compared to
 the
transverse waveguide frequency $\omega _{\perp }$. Due to coupling to
 excited transverse
states, the interaction strength demonstrates a resonant dependence on
the ratio of the elastic scattering length $a_{3D}$  to the transverse
oscillator length
\begin{equation}
a_{\perp }=\sqrt{{2\over m\omega { } _{\perp }}} ,
\end{equation}
\medskip
where $m$ is the atomic mass. This confinement-induced resonance can
been interpreted as a Feshbach resonance (see Refs.\
\cite{BMO03,GNO04}), where the excited transverse states play the role
of a closed channel. The interaction strength demonstrates also a
dependence on the collision energy, which is a common property of a
resonant scattering. Such an energy dependence appears in the
 scattering
amplitudes in Ref.\ \cite{MBO04}, as well as in the bound state energy
in Ref.\ \cite{BMO03}. However, the use of Feshbach resonance for
 tuning
the elastic scattering length, as in Ref.\ \cite{MSGKE05}, leads to
additional energy dependence (see Ref.\ \cite{Y05}), which can be
important for a weak resonance. A similar effect has been considered
 in
Refs.\ \cite{TWMJ00,BG02,BTJ02,M05} for a problem of two atoms under
 3D
harmonic confinement.

A Feshbach resonance appears when the collision energy of an atomic
pair in the open channel lies in a vicinity of a bound (molecular)
 state
in a closed channel (see Ref.\ \cite{TTHK99}). As a result, the
 quasi-1D
molecules are superpositions of the closed channel state and the
 ground
and excited waveguide modes of the open channel. The closed-channel
contribution becomes substantial for weak resonances.

The theory of two-body two-channel problem under tight cylindrical
harmonic confinement is summarized in Sec.\ \ref{SecFeshWG}. This
problem can be approximated by a 1D two-channel model. An improved
relation between the 1D and 3D scattering parameters, presented in
 Sec.\
\ref{SecRel1D} below, increases the range of applicability of the 1D
model compared to the relations in Ref.\ \cite{Y05}. The composition
 of
quasi-1D bound states is analyzed in Sec.\ \ref{BoundStComp}.
 Three-body
1D molecules are considered in Sec.\ \ref{ThreeAtom} in a way which is
similar to the analysis of scattering in Ref.\ \cite{YBO06}. A system
 of
units with $\hbar =1$ is used below.

\section{Feshbach resonance in harmonic waveguides\label{SecFeshWG}}

The properties of two-atom systems can be described by
close-coupled equations for the wavefunction of the open channel
$\psi _{a}\left( {\bf r}\right) $ and the amplitude for the system to
 be in the closed
channel $\psi _{m}$,  of the form (see Ref.\ \cite{Y05}),
\begin{eqnarray}
E\psi _{a}\left( {\bf r}\right) &=&\left\lbrack -{1\over m}\nabla
 ^{2}+V_{a}\delta \left( {\bf r}\right) +V_{\text{conf}}\left( {\bf
 r}\right) \right\rbrack \psi _{a}\left( {\bf r}\right)  \nonumber
\\
&&+V_{am}\delta \left( {\bf r}\right) \psi _{m}\label{CC3D}
\\
E\psi _{m}&=&D_{3D}\psi _{m}+V^{*}_{am}\psi _{a}\left( 0\right)  .
 \nonumber
\end{eqnarray}
Here $E$ and ${\bf r}$ are, respectively, the energy and coordinate
 vector of
the relative motion and all the energies are measured from the open
 channel
threshold. For a harmonic waveguide the confinement potential has the
 form
\begin{equation}
V_{\text{conf}}={m\over 4}\omega ^{2}_{\perp }\rho ^{2} ,
\end{equation}
where $\rho $ and $z$ are the cylindrical components of the vector
 ${\bf r}$. The
strength of the open channel potential $V_{a}$, the coupling strength
 $V_{am}$, and
the bound state energy in the closed channel $D_{3D}$  can be
 expressed as (see
Ref.\ \cite{Y05})
\begin{eqnarray}
&&V_{a}={4\pi \over m}a_{3D}\left( 1-{2\over \pi }a_{3D}p_{c}\right)
 ^{-1} \nonumber
\\
&&|V_{am}|^{2}={4\pi \over m}a_{3D}\mu \Delta \left( 1-{2\over \pi
 }a_{3D}p_{c}\right) ^{-2} \label{renorm}
\\
&&D_{3D}=\mu \left\lbrack B-B_{0}-\Delta +\Delta \left( 1-{2\over \pi
 }a_{3D}p_{c}\right) ^{-1}\right\rbrack  \nonumber
\end{eqnarray}
in terms of the phenomenological resonance strength $\Delta $, the
 difference
between the magnetic momenta of an atomic pair in the open and closed
channels $\mu $, the detuning of the external magnetic field $B$ from
 its resonant
value $B_{0}$, and the momentum cutoff $p_{c}$. The final results
 reached below are
derived in the limit $p_{c}\rightarrow \infty $. The $\delta
 $-function potentials are applicable to two
indistinguishable bosons, as well as to bosons or fermions with
 different
spins.

Elimination of $\psi _{m}$  from Eqs.\ (\ref{CC3D}) leads to a single
 equation
for $\psi _{a}\left( {\bf r}\right) $. It can be expanded in terms of
 the transverse Hamiltonian
eigenfunctions $|n0\rangle $ with the zeroth angular momentum
 projection on the
waveguide axis $z$ as
\begin{equation}
\psi _{a}\left( {\bf r}\right) =\left( 2\pi \right) ^{-1
/2}\sum\limits^{\infty }_{n=0}\int\limits^{\infty }_{-\infty }dq
 \tilde{\psi }_{n}\left( q\right)  e^{iqz}|n0\rangle  . \label{psia}
\end{equation}
The coefficients $\tilde{\psi }_{n}\left( q\right) $ satisfy the set
 of coupled equations (see Ref.\
\cite{Y05})
\begin{equation}
{p^{2}_{n}-q{ } ^{2}\over m}\tilde{\psi }_{n}\left( q\right) ={1\over
 2\pi ^{2}a{ } ^{2}_{\perp }}V_{\text{eff}}\left( E\right)
 \sum\limits^{\infty }_{n^\prime =0}\int\limits^{\infty }_{-\infty
 }dq^\prime \tilde{\psi }_{n^\prime }\left( q^\prime \right)  ,
 \label{phin}
\end{equation}
where
\begin{equation}
V_{\text{eff}}\left( E\right) =V_{a}+{|V_{am}|{ } ^{2}\over E-D{ }
 _{3D}} \label{Veff}
\end{equation}
is a non-renormalized energy-dependent interaction strength and
\begin{equation}
p_{n}=\sqrt{m\left\lbrack E-\left( 2n+1\right) \omega _{\perp
 }\right\rbrack } \label{pn}
\end{equation}
is the relative axial momentum for the channel corresponding to the
transverse excited state $|n0\rangle $ with the excitation energy
 $\left( 2n+1\right) \omega _{\perp }$.

The transition matrix for a two-atom collision in an atomic waveguide
has been derived in Ref.\ \cite{Y05} as
\begin{equation}
T_{\text{conf}}\left( p_{0}\right) ={4\over ma{ } _{\perp
 }}\left\lbrack {a{ } _{\perp }\over a_{\text{eff}}\left( E\right) }
+\zeta \left( {1\over 2},-\left( {a_{\perp }p{ } _{0}\over 2}\right)
 ^{2}\right) \right\rbrack ^{-1} . \label{Tconf}
\end{equation}
Although the collision momentum $p_{n}$  depends on the channel, the
transition matrix is independent of the initial and final transverse
 states
$n$, $n^\prime $ for all open channels ($n, n^\prime <\left( E/\omega
 _{\perp }-1\right) /2$) and is expressed by Eq.\
(\ref{Tconf}) in terms of $p_{0}$. It is a consequence of the use of
 zero-range
potentials in Eqs.\ (\ref{CC3D}). The energy-dependent length
\begin{equation}
a_{\text{eff}}\left( E\right) =a_{3D}\left\lbrack 1+{\mu \Delta \over
 E-\mu \left( B-B_{0}\right) }\right\rbrack
\end{equation}
replaces the elastic scattering length in Bethe-Peierls boundary
condition. The Hurwitz zeta function is defined as (see Refs.\
\cite{MBO04,BE53}),
\begin{equation}
\zeta \left( \nu ,\alpha \right) =\mathrel{\mathop
 \mathrm{\lim}_{n_{c}\rightarrow \infty }}\left\lbrack
 \sum\limits^{n{ } _{c}}_{n=0}\left( n+\alpha \right) ^{-\nu
 }-{1\over 1-\nu }\left( n_{c}+\alpha \right) ^{1-\nu }\right\rbrack
 , \label{zeta}
\end{equation}
with $-2\pi <\arg( n+\alpha ) \le 0$.

\section{Relation to the one-dimensional problem\label{SecRel1D}}

The confined two-body problem can be interpreted as a 1D one described
by the Schr\"odinger equation
\begin{equation}
E_{c}\varphi _{0}\left( z\right) =-{1\over m} {d^{2}\varphi { }
 _{0}\over dz{ } ^{2}}+U_{\text{eff}}\left( E_{c}\right) \delta \left
( z\right) \varphi _{0}\left( 0\right)  \label{Effam}
\end{equation}
with a zero-range interaction, where the interaction strength
 $U_{\text{eff}}$
depends on the collision energy $E_{c}=p^{2}_{0}/m$. The 1D
 transition matrix
corresponding to Eq.\ (\ref{Effam}),
\begin{equation}
T_{1D}\left( p_{0}\right) =\left\lbrack U^{-1}_{\text{eff}}\left(
 E_{c}\right) +{im\over 2p{ } _{0}}\right\rbrack ^{-1} , \label{T1D}
\end{equation}
coincides with Eq.\ (\ref{Tconf}) for
\begin{equation}
U_{\text{eff}}\left( E_{c}\right) =\left\lbrack {1\over 2\omega
 _{\perp }a_{\text{eff}}\left( E_{c}+\omega _{\perp }\right) }+{ma{ }
 _{\perp }\over 4}\zeta \left( {1\over 2},-{E{ } _{c}\over 2\omega {
 } _{\perp }}\right) -{im\over 2p{ } _{0}}\right\rbrack ^{-1} .
 \label{Ueff}
\end{equation}
The case of low collision energies $E_{c}\ll 2\omega _{\perp }$  can
 be analyzed using
the expansion (see Refs.\ \cite{O98,MBO04})
\begin{equation}
\zeta \left( {1\over 2},\alpha \right) \mathrel{\mathop \sim_{\alpha
 \rightarrow 0}}{1\over \sqrt{\alpha }}-C-C^\prime \alpha , \sqrt{-
|\alpha |}=-i\sqrt{|\alpha |}, \label{Zeta3}
\end{equation}
where $C=-\zeta \left( {1\over 2}\right) \approx 1.4603$, $C^\prime
 ={1\over 2}\zeta \left( {3\over 2}\right) \approx 1.3062$, and
 $\zeta \left( \nu \right) $ is the Riemann
$\zeta $-function (see Ref.\ \cite{BE53}).

Following expressions attain a simpler form written in terms of
dimensionless parameters (the scattering momentum $k$, the elastic
scattering length $a$, the detuning $b^\prime $, and the resonance
 strength $d$),
defined as
\begin{eqnarray}
k={p_{0}a{ } _{\perp }\over 2} ,\qquad a={a{ } _{3D}\over a{ }
 _{\perp }} \nonumber
\\
b^\prime =\mu {B-B_{0}-\Delta \over 2\omega { } _{\perp }}-{1\over 2}
 ,\qquad d={a_{3D}\mu \Delta \over 2a_{\perp }\omega { } _{\perp }} .
 \label{kabd}
\end{eqnarray}
The parameters $k$, $a$, and $d$ have been used previously in Ref.\
\cite{Y05}. The detuning $b^\prime $, measured from the crossing
 point of the
closed-channel bound state and the continuum threshold
($D_{3D}=\omega _{\perp }$), is
related to the detuning $b$ of Ref.\ \cite{Y05}, measured from the
scattering length resonance, as $b^\prime =b-d/a$.

Substitution of the expansion (\ref{Zeta3}) into Eq.\
(\ref{Ueff}) leads to the expression of the interaction strength in
the form
\begin{equation}
U_{\text{eff}}\left( 2\omega _{\perp }k^{2}\right) ={4\over ma{ }
 _{\perp }} a {k^{2}-b^\prime \over C^\prime ak^{4}+\beta k^{2}-\left
( 1-Ca\right) b^\prime -d/a} , \label{Ueff1}
\end{equation}
where
\begin{equation}
\beta =1-Ca-C^\prime ab^\prime  .
\end{equation}
In a wide range of the parameters the interaction strength can be
approximately expressed as
\begin{equation}
U_{\text{eff}}\left( E_{c}\right) =U_{a}+{2|g|{ } ^{2}\over E_{c}-D{
 } _{1D}} . \label{Ueff1D}
\end{equation}
This form corresponds to a two-channel 1D problem (see Ref.\
\cite{Y05}), described by the coupled equations
\begin{eqnarray}
&&E\varphi _{0}\left( z\right) =-{1\over m} {d^{2}\varphi { }
 _{0}\over dz{ } ^{2}}+\left\lbrack U_{a}\varphi _{0}\left( 0\right)
+\sqrt{2}g^{*}\varphi ^{am}_{1}\right\rbrack \delta \left( z\right)
 \nonumber
\\
&&{}\label{CC1D}
\\
&&E\varphi ^{am}_{1}=D_{1D}\varphi ^{am}_{1}+\sqrt{2}g\varphi
 _{0}\left( 0\right)  \nonumber
\end{eqnarray}
for the open- and closed-channel coefficients $\varphi _{0}\left(
 z\right) $ and $\varphi ^{am}_{1}$,
respectively. The non-resonant interaction strength $U_{a}$, the
 channel
coupling $g$, and the detuning $D_{1D}$  will be further related to
 the 3D
scattering parameters. Equation (\ref{Effam}) with $U_{\text{eff}}$
 given by Eq.\
(\ref{Ueff1D}) can be obtained by the elimination of the closed
 channel
from Eqs.\ (\ref{CC1D}).

The ratio of the first term in the denominator of Eq.\
(\ref{Ueff1}) to the other terms does not exceed the order of
 magnitude
of $ak^{2}$. It can be neglected whenever $a<1$ and $k\ll 1$, leading
 to an
expression of the form of Eq.\ (\ref{Ueff1D}) with
\begin{eqnarray}
U_{a}={4a\over ma_{\perp }\beta } ,\qquad |g|^{2}=4\omega _{\perp }{d
+C^\prime \left( ab^\prime \right) { } ^{2}\over ma_{\perp }\beta { }
 ^{2}} \nonumber
\\
D_{1D}=2\omega _{\perp }{\left( 1-Ca\right) ab^\prime +d\over a\beta
 } . \label{UgD}
\end{eqnarray}
For the case of a relatively small detuning, $C^\prime |ab^\prime
|\ll |1-Ca|$, or
\begin{equation}
|\mu \left( B-B_{0}-\Delta -\omega _{\perp }/\mu \right) |\ll
|{a_{\perp }\omega { } _{\perp }\over a{ } _{3D}}\left( 1-C{a{ }
 _{3D}\over a{ } _{\perp }}\right) | , \label{dbll}
\end{equation}
the terms proportional to $b^\prime $  in the parameter $\beta $ can
 be
neglected. The parameters $U_{a}$  and $D_{1D}$  are expressed then
 by Eqs.\
(43) and (45) in Ref.\ \cite{Y05}, while Eq.\ (44) therein will be
valid whenever $d\gg a^{2}b^\prime { } ^{2}$, or $\mu ^{2}\left(
 B-B_{0}-\Delta -\omega _{\perp }/\mu \right) ^{2}\ll a_{\perp
 }\omega _{\perp }|\mu \Delta /a_{3D}|$.

\begin{figure}
\includegraphics[width=3.375in]{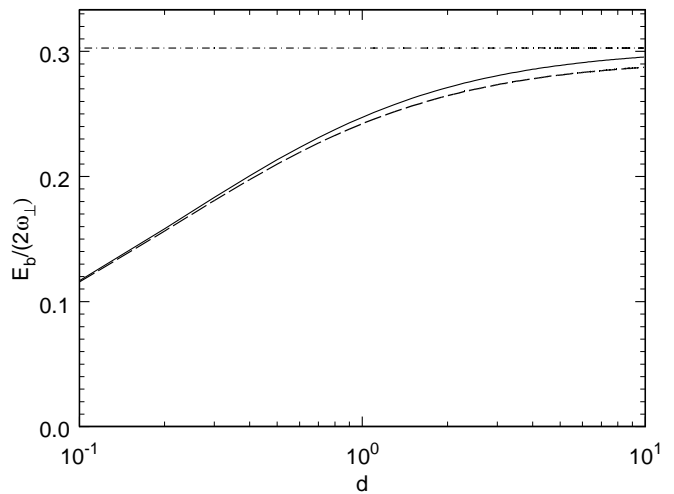}

\caption{The binding energy $E_{b}$  calculated as a function of the
dimensionless resonance strength $d$ [see Eq.\ (\protect\ref{kabd})]
 at
the scattering length resonance $B=B_{0}+\omega _{\perp }/\mu $, or
 $b^\prime =-d/a$, with the
dimensionless non-resonant interaction strength $a=0.1$. The solid,
dashed, and dot-dashed lines present, respectively, solutions of the
exact equation  (\protect\ref{x}), the 1D approximation
(\protect\ref{x1D}), and the open-channel model (\protect\ref{xOC}).}
\label{FigEbd}

\end{figure}

Unlike Eqs.\ (43)-(45) in Ref.\ \cite{Y05}, the relations
(\ref{UgD}) demonstrate a non-linear dependence of all three
parameters $U_{a}$, $g$, and $D_{1D}$  on the elastic scattering
 length, detuning,
and resonance strength. These relations substantially increase the
applicability range of the 1D approximation, as is demonstrated by
Fig.\ \ref{FigEbd} using the example of binding energy $E_{b}=\omega
 _{\perp }-E=2\omega _{\perp }x^{2}$.
The parameter $x$ here is the ratio of $a_{\perp }$  to the bound
 state axial
size. For the {\it confined system} it is determined as a solution of
 the
transcendent equation (see Eq.\ (55) in Ref.\ \cite{Y05})
\begin{equation}
{ax^{2}+ab^\prime +d\over x^{2}+b^\prime }=-a^{2}\zeta \left( {1\over
 2},x^{2}\right) , x>0, \label{x} .
\end{equation}
while for the related 1D system it is evaluated as a solution of
the cubic equation
\begin{equation}
\beta x^{3}+ax^{2}+\left\lbrack \left( 1-Ca\right) b^\prime +d
/a\right\rbrack x+ab^\prime =0 . \label{x1D}
\end{equation}
This equation determines poles of the 1D $T$-matrix (\ref{T1D})
with the interaction strength (\ref{Ueff1D}) and the resonance
parameters given by Eq.\ (\ref{UgD}). A similar equation has been
considered in Ref.\ \cite{KD98}.

Substitution of the resonant scattering length $a_{3D}\left\lbrack
 1-\Delta /\left( B-B_{0}-\omega _{\perp }/\mu \right) \right\rbrack
 $,
which takes into account the resonance shift by $\omega _{\perp }/\mu
 $ due to confinement,
into the equations of Ref.\ \cite{BMO03}, leads to the following
 equation
for the parameter $x$
\begin{equation}
{ab^\prime +d\over b^\prime }=-a^{2}\zeta \left( {1\over
 2},x^{2}\right) , x>0, \label{xOC} .
\end{equation}
This approximation, corresponding to a single-channel confined
problem with an energy-independent interaction, is called here the
``{\it open-channel model}''. Figure \ref{FigEbd} demonstrates that
 this
model is applicable to strong resonances only.

\begin{figure}
\includegraphics[width=3.375in]{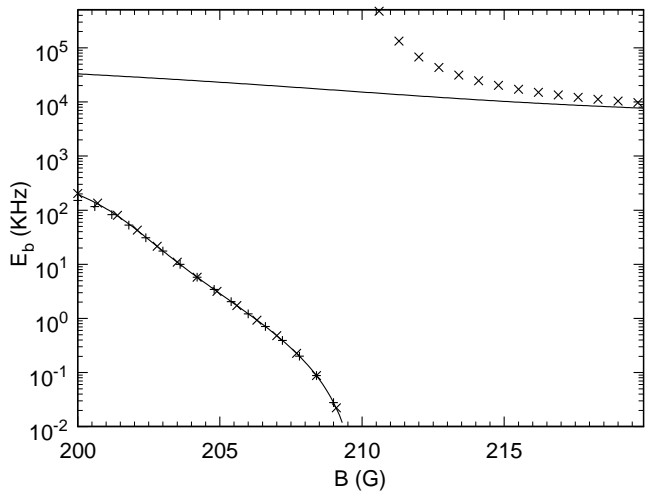}

\caption{The binding energy $E_{b}$  calculated as a function of the
 external
magnetic field for the 202 G resonance in K with $\omega _{\perp
 }=69\times 2\pi $  KHz. The solid
line, pluses, and crosses present, respectively, solutions of the
 exact
equation  (\protect\ref{x}), the 1D approximation (\protect\ref{x1D}
), and
the open-channel model  (\protect\ref{xOC}).} \label{FigEbK}

\end{figure}

Various approximations for the binding energy are compared in Fig.\
\ref{FigEbK} for a strong resonance. The results demonstrate good
agreement between solutions of the exact equation (\ref{x}), the 1D
approximation (\ref{x1D}), and the open-channel model  (\ref{xOC}) for
$B<B_{0}+\Delta +\omega _{\perp }/\mu $, when a weak bound state
 exists. However, the open-channel
model predicts a non-physical singularity at $B=B_{0}+\Delta +\omega
 _{\perp }/\mu $. Higher above the
resonance the result of the open-channel model tends to the energy of
 the
deep quasi-3D bound state. The latter state is not described by the 1D
approximation. The exact equation (\ref{x}) gives correctly both the
 deep
and weak bound states.

\begin{figure}
\includegraphics[width=3.375in]{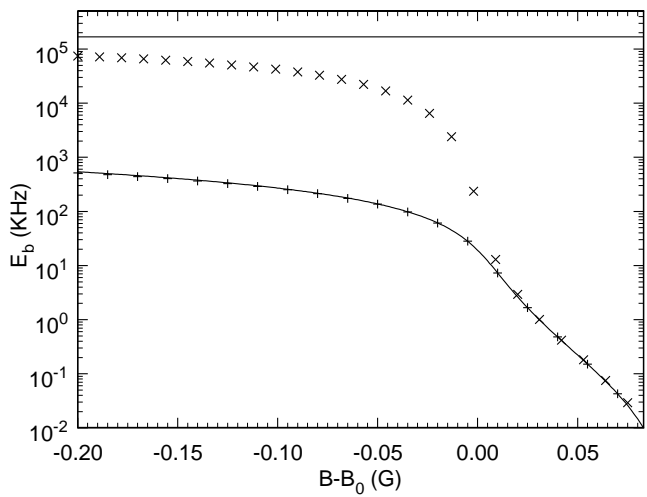}

\caption{The binding energy $E_{b}$  calculated as a function of the
external magnetic field for the 543 G resonance in $^{6}$Li with
 $\omega _{\perp }=200\times 2\pi $  KHz.
The solid line, pluses, and crosses present, respectively, solutions
 of the
exact equation  (\protect\ref{x}), the 1D approximation
(\protect\ref{x1D}),
and the open-channel model  (\protect\ref{xOC}).} \label{FigEbLi}

\end{figure}

The 1D approximation agrees to the exact equations also in the case of
a weak resonance (see Fig.\ \ref{FigEbLi}). However the open-channel
 model,
neglecting energy dependence of  the interaction strength, is
 applicable in
this case only within a small interval of $B$ close to $B_{0}+\Delta
+\omega _{\perp }/\mu $.

The applicability range of the 1D approximation is improved
because the approximation (\ref{Zeta3}) has a root at $\alpha
 =0.294$, close
to exact value of 0.303 (cf. the value of 0.468 provided by two-term
approximation used in Ref.\ \cite{Y05}). The accuracy of the binding
energy demonstrates the applicability of the 1D approximation to the
off-shell $T$-matrix, which is important for applications to many-body
problems (see also discussion in Ref.\ \cite{AZ85}). The only criteria
of applicability are $a<1$ and $E_{b}\ll \omega _{\perp }$  (or
 $E_{c}\ll \omega _{\perp }$  for collisions).

Equation (\ref{Ueff1}) can demonstrate an energy-dependence of
the form of Eq.\ (\ref{Ueff1D}) for a weak resonance, whenever $d\ll
 ab^\prime $,
or
\begin{equation}
\Delta \ll |B-B_{0}-\Delta -\omega _{\perp }/\mu |
\end{equation}
and the term $d/a$ in the denominator of Eq.\ (\ref{Ueff1}) can be
neglected. In this case the 1D parameters can be estimated as
\begin{equation}
U_{a}\approx 0, |g|^{2}\approx {4\omega { } _{\perp }\over C^\prime
 ma{ } _{\perp }}, D_{1D}\approx -2\omega _{\perp }{1-Ca\over
 C^\prime a} . \label{UgDcir}
\end{equation}
They are  independent of the Feshbach resonance parameters. Therefore,
in agreement with Ref.\ \cite{BMO03}, a confinement-induced resonance
 can be
interpreted as a two-state Feshbach resonance even for non-resonant 3D
scattering. It is a consequence of the approximation (\ref{Zeta3}). It
should be noted that the closed channel involves in this case a
superposition of all transverse excitations. However, the detuning
 $D_{1D}$  then
substantially exceeds the transverse frequency $\omega _{\perp }$
 and the energy-dependence
is very weak in the quasi-1D regime, whenever the energy is less then
 $\omega _{\perp }$.

\section{Bound state composition \label{BoundStComp}}

The bound states of two atoms in atomic waveguides are
superpositions of the closed and open channels of the Feshbach
resonance. The open-channel component is a superposition of all
transverse modes. The size of the closed-channel component is
negligibly small compared to $a_{\perp }$, this component is not
 affected by the
waveguide, and does not need an expansion in terms of the transverse
modes.

For a bound state with energy $E<\omega _{\perp }$  all momenta
 $p_{n}$,  defined by
Eq.\ (\ref{pn}), are imaginary and the solution of Eq.\ (\ref{phin})
has the form
\begin{equation}
\tilde{\psi }_{n}\left( q\right) ={C{ } _{a}\over |p_{n}|^{2}+q{ }
 ^{2}} .
\end{equation}
The probability to find the molecule in the $n$-th transverse mode
of the open channel can then be expressed as
\begin{equation}
W_{n}={\pi \over 2} {|C_{a}|{ } ^{2}\over |p_{n}|{ } ^{3}} .
\end{equation}
Equations (\ref{CC3D}) and (\ref{psia}) allow to relate the
closed-channel amplitude $\psi _{m}$  to the open-channel
 wavefunction as
\begin{equation}
\psi _{m}={V{ } ^{*}_{am}\over E-D{ } _{3D}} {C{ } _{a}\over
 \sqrt{2}\pi a{ } _{\perp }}\sum\limits^{\infty
 }_{n=0}\int\limits^{\infty }_{-\infty }\tilde{\psi }_{n}\left(
 q\right) dq .
\end{equation}
Although the sum here diverges, the renormalization procedure of
Ref.\ \cite{Y05} leads to a finite probability to find the molecule in
the closed channel
\begin{equation}
W_{c}=|\psi _{m}|^{2}={\pi |C_{a}|^{2}a^{2}_{\perp }\mu \Delta \over
 2ma_{3D}\left\lbrack E-\mu \left( B-B_{0}-\Delta \right)
 \right\rbrack { } ^{2}} .
\end{equation}
The final results can be expressed in terms of the dimensionless
parameters (\ref{kabd})
\begin{equation}
{W{ } _{c}\over W{ } _{0}}={2dx{ } ^{3}\over a^{2}\left( x^{2}
+b^\prime \right) { } ^{2}} ,\qquad {W{ } _{e}\over W{ }
 _{0}}=x^{3}\zeta \left( {3\over 2},x^{2}\right) -1 ,
\end{equation}
where $x$ is the solution of Eq.\ (\ref{x}),
 $W_{e}=\sum\limits^{\infty }_{n=1}W_{n}$  is the total
contribution of all excited transverse states, and the contributions
are normalized as $W_{0}+W_{e}+W_{c}=1$. An approximate expression,
 used in
following calculations,
\begin{eqnarray}
\zeta \left( {3\over 2},x^{2}\right) \approx {1+2x{ } ^{2}\over x{ }
 ^{3}} \nonumber
\\
\times \left\lbrack 1-{x{ } ^{2}\over 0.5+0.54884x+2.5636x^{2}
+0.12172x^{3}+4x{ } ^{4}}\right\rbrack  \nonumber
\end{eqnarray}
gives a relative error of less then $3\times 10^{-4}$.

Bound states of the related 1D system are superpositions of open
and closed channels, where the closed channel effectively incorporates
contributions of the excited transverse states and the closed channel
of the confined system. The two contributions can be respectively
expressed as
\begin{equation}
W^{1D}_{c}=|\varphi ^{am}_{1}|^{2} ,\qquad
 W^{1D}_{o}=\int\limits^{\infty }_{-\infty }|\varphi _{0}\left(
 z\right) |^{2}dz ,
\end{equation}
where the bound-state solution of Eqs.\ (\ref{CC1D}) has the form
\begin{equation}
\varphi ^{am}_{1}= {\sqrt{2}g\varphi _{0}\left( 0\right) \over E-D{ }
 _{1D}} , \varphi _{0}\left( z\right) =\varphi _{0}\left( 0\right)
 \exp\left( -\sqrt{mE{ } _{b}}|z|\right)  .
\end{equation}
The ratio of the contributions can be expressed as
\begin{equation}
{W{ } ^{1D}_{c}\over W{ } ^{1D}_{o}}=2x^{3}{d+C^\prime a^{2}b^\prime
 { } ^{2}\over a^{2}\left( x^{2}+b^\prime \right) { } ^{2}} .
\end{equation}
A direct evaluation demonstrates that
\begin{equation}
{W_{c}+W{ } _{e}\over W{ } _{0}}\mathrel{\mathop \sim_{x\rightarrow
 0}}{W{ } ^{1D}_{c}\over W{ } ^{1D}_{o}}\text{  , where }\zeta \left(
 {3\over 2},x^{2}\right) \mathrel{\mathop \sim_{x\rightarrow
 0}}x^{-3}+\zeta \left( {3\over 2}\right) .
\end{equation}
\begin{figure}
\includegraphics[width=3.375in]{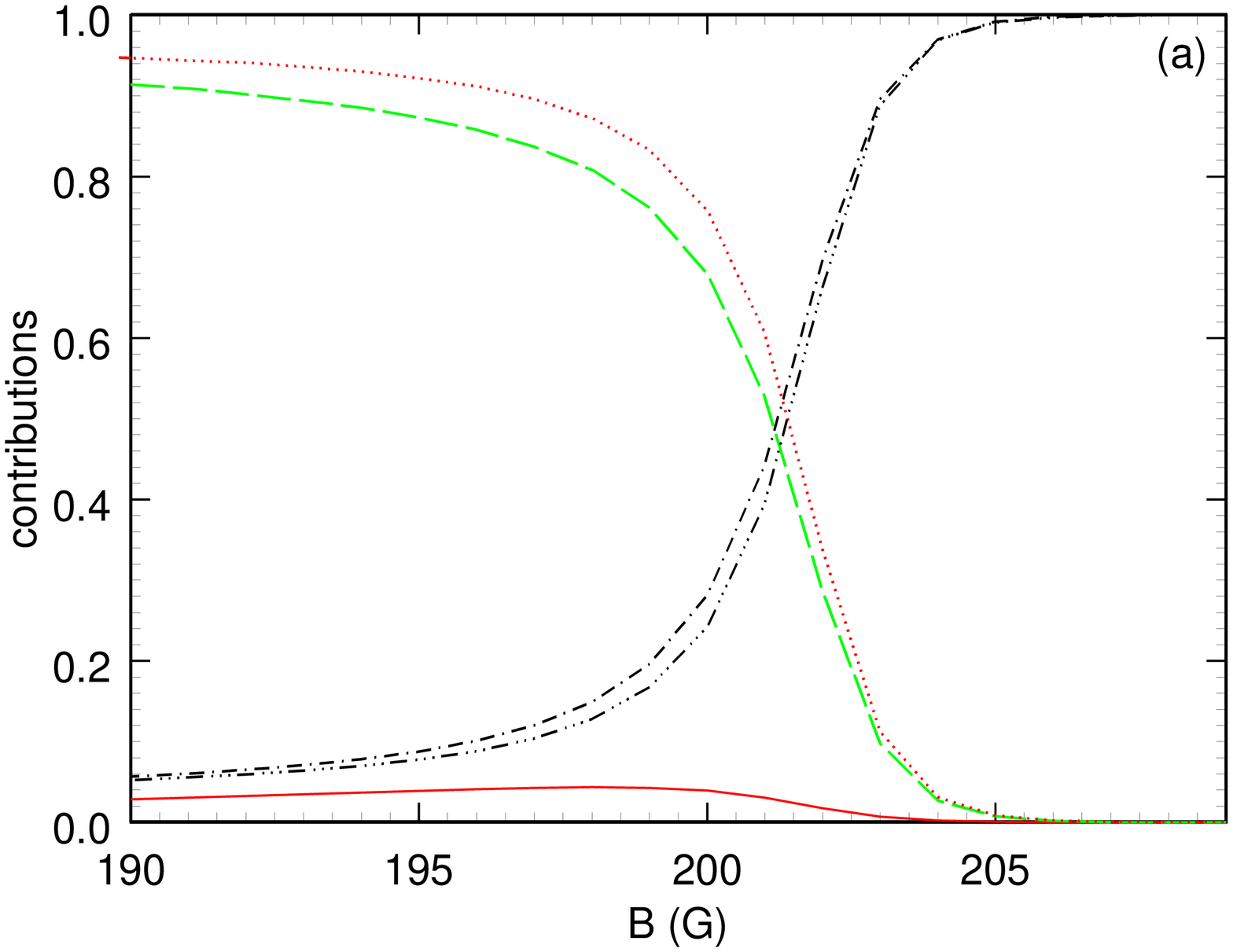}
\includegraphics[width=3.375in]{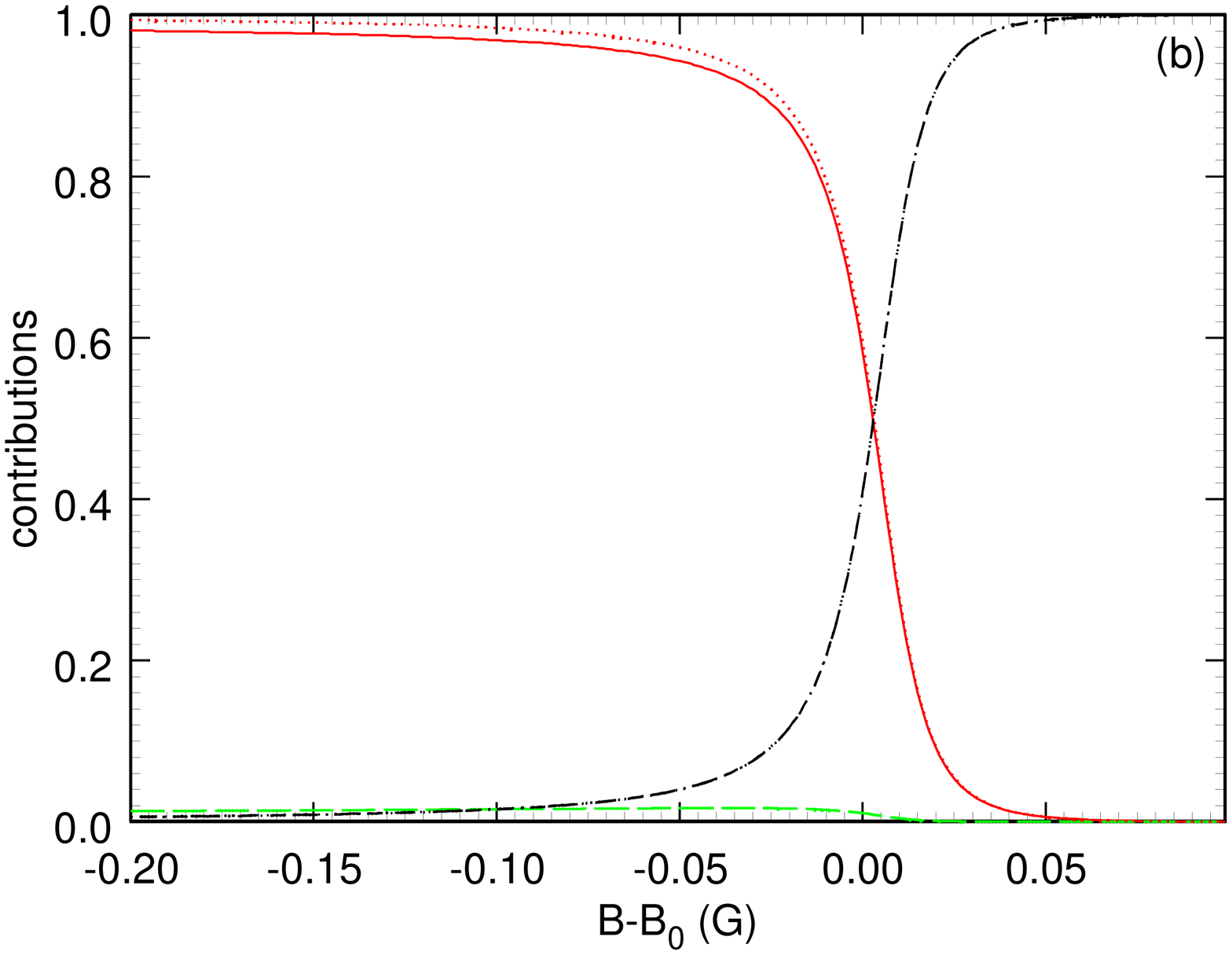}

\caption{(color online) The contributions of the ground (dot-dashed
line) and excited (dashed line) states of the open channel and of the
 closed
channel (solid line) to the weak bound state calculated as a function
 of the
external magnetic field. The dot-dot-dot-dashed and dotted lines
 present,
respectively, contributions of the open and closed channels in the 1D
 model.
The parts (a) and (b) are related, respectively, to the 202 G
 resonance in K
with $\omega _{\perp }=69\times 2\pi $  KHz and to the 543 G
 resonance in $^{6}$Li with $\omega _{\perp }=200\times 2\pi $  KHz.
The dot-dashed and dot-dot-dot-dashed lines almost coincide in the
 part
(b).} \label{FigBScomp}

\end{figure}

The contributions to the bound states are presented in Fig.\
\ref{FigBScomp} for two resonances: the strong one in K with
 $ad\approx 1.6$
and the weak one in $^{6}$Li with $ad\approx 4.1\times 10^{-4}$. The
 results demonstrate that
$W^{1D}_{c}\approx W_{e}+W_{c}$  and $W^{1D}_{o}\approx W_{0}$  with
 a good accuracy over a wide range of
parameter values. The closed channel and excited states of the open
channel yield the dominant contribution far below the bound state
threshold at $B=B_{0}+\Delta +\omega _{\perp }/\mu $, while the
 ground state of the open channel
becomes dominant near the threshold. For the strong resonance the
contribution of the closed channel in the confined system is always
small, and the bound state consists mostly of the ground and excited
states of the open channel. However, for the weak resonance the major
contribution is yielded by the closed channel and the ground state of
the excited channel. The closed channel and excited states of the open
channel yield similar contributions far below the bound state
threshold for $ad=ma^{2}_{\text{3D}}\mu \Delta /4\approx 0.1$. The
 resonances with higher or lower
values of the product $ad$ can be called, respectively, as
 open-channel
or closed-channel dominated resonances, as in a case of free space
(see Refs.\ \cite{GKGTJ04,Chin05}).

\section{Three-atom bound states\label{ThreeAtom}}

The previous results demonstrate that two atoms with a resonant
interaction in an atomic waveguide can be within a good accuracy
considered as 1D particles with a resonant interaction. Consider now a
three-body 1D problem for bosonic atoms. (The same approach has been
used in Ref.\ \cite{YBO06} for the analysis of three-body scattering.)
A state vector can be represented in the form
\begin{eqnarray}
|\Psi _{3}\rangle =\biggl\lbrack \int dzdz_{m}\varphi ^{\left(
 3\right) }_{1}\left( z,z_{m}\right) \hat{\Psi }^{\dag }_{a}\left(
 z\right) \hat{\Psi }^{\dag }_{m}\left( z_{m}\right)  \nonumber
\\
+{1\over \sqrt{6}}\int d^{3}z\varphi ^{\left( 3\right) }_{0}\left(
 z_{1},z_{2},z_{3}\right) \hat{\Psi }^{\dag }_{a}\left( z_{1}\right)
 \hat{\Psi }^{\dag }_{a}\left( z_{2}\right) \hat{\Psi }^{\dag
 }_{a}\left( z_{3}\right) \biggr\rbrack |0\rangle  \label{Psi3}
\end{eqnarray}
as a superposition of the three-atom channel, described by the
coefficient $\varphi ^{\left( 3\right) }_{0}\left(
 z_{1},z_{2},z_{3}\right) $, and the atom-molecule channel (involving
the closed-channel molecules), described by the coefficient
$\varphi ^{\left( 3\right) }_{1}\left( z,z_{m}\right) $. Here
 $\hat{\Psi }^{\dag }_{a}\left( z\right) $ and $\hat{\Psi }^{\dag
 }_{m}\left( z_{m}\right) $ are the creation operators for the
atom and closed-channel molecule, respectively, and $z$, $z_{m}$  are
 their
coordinates. Substitution of Eq.\ (\ref{Psi3}) into the Schr\"odinger
equation with the Hamiltonian (25) of Ref.\ \cite{Y05}  leads to the
following coupled equations
\begin{eqnarray}
E\varphi ^{\left( 3\right) }_{0}\left( z_{1},z_{2},z_{3}\right)
 =\biggl\lbrack -{1\over 2m}\sum\limits^{3}_{j=1}{\partial { }
 ^{2}\over \partial z{ } ^{2}_{j}}+U_{a}\lbrack \delta \left(
 z_{1}-z_{2}\right)  \nonumber
\\
+\delta \left( z_{2}-z_{3}\right) +\delta \left( z_{1}-z_{3}\right)
 \rbrack  \biggr\rbrack \varphi ^{\left( 3\right) }_{0}\left(
 z_{1},z_{2},z_{3}\right)  \nonumber
\\
+\sqrt{{2\over 3}}g^{*}\bigl\lbrack \varphi ^{\left( 3\right)
 }_{1}\left( z_{1},z_{2}\right) \delta \left( z_{2}-z_{3}\right)
 \nonumber
\\
+\varphi ^{\left( 3\right) }_{1}\left( z_{2},z_{1}\right) \delta
 \left( z_{1}-z_{3}\right) +\varphi ^{\left( 3\right) }_{1}\left(
 z_{3},z_{1}\right) \delta \left( z_{1}-z_{2}\right) \bigr\rbrack
 \label{CC3B}
\\
E\varphi ^{\left( 3\right) }_{1}\left( z,z_{m}\right) =\left\lbrack
 -{1\over 2m}{\partial { } ^{2}\over \partial z{ } ^{2}}-{1\over
 4m}{\partial { } ^{2}\over \partial z{ } ^{2}_{m}}
+D_{1D}\right\rbrack \varphi ^{\left( 3\right) }_{1}\left(
 z,z_{m}\right)  \nonumber
\\
+\sqrt{6}g\varphi ^{\left( 3\right) }_{0}\left( z,z_{m},z_{m}\right)
 \nonumber
\end{eqnarray}
The indistinguishability of the bosonic atoms leads to the symmetry of
the wavefunction $\varphi ^{\left( 3\right) }_{0}\left(
 z_{1},z_{2},z_{3}\right) $ over permutation of the atomic
 coordinates.

Equations (\ref{CC3B}), as well as the Hamiltonian, do not contain
terms describing collisions between the closed-channel molecule and
 the
third atom. This assumption is justified since the atoms in the
 closed and
open channels have different hyperfine states. It allows a simple
elimination of the atom-molecule channel. The analysis of the
 remaining
three-atom channel attains a simpler form in the momentum
 representation.
The corresponding three-atom wavefunction
\begin{eqnarray}
\tilde{\varphi }^{\left( 3\right) }_{0}\left( q_{1},q_{3},q_{3}\right
) =\left( 2\pi \right) ^{-3/2}\int d^{3}q \exp\left(
 -i\sum\limits^{3}_{j=1}q_{j}z_{j}\right)  \nonumber
\\
\times \varphi ^{\left( 3\right) }_{0}\left( z_{1},z_{2},z_{3}\right)
\end{eqnarray}
obeys the single-channel Schr\"odinger equation
\begin{eqnarray}
E \tilde{\varphi }^{\left( 3\right) }_{0}\left(
 q_{1},q_{2},q_{3}\right) ={1\over
 2m}\sum\limits^{3}_{j=1}q^{2}_{j}\tilde{\varphi }^{\left( 3\right)
 }_{0}\left( q_{1},q_{2},q_{3}\right)  \nonumber
\\
+{1\over 2\pi }\sum\limits^{3}_{j=1}U_{\text{eff}}\left( E-{1\over
 2m}q^{2}_{j}-{1\over 4m}\left( Q-q_{j}\right) ^{2}\right)  \nonumber
\\
\times \int d^{3}q^\prime \delta \left( q^\prime _{j}-q_{j}\right)
 \delta \left( Q-Q^\prime \right) \tilde{\varphi }^{\left( 3\right)
 }_{0}\left( q_{1}^\prime ,q_{2}^\prime ,q_{3}^\prime \right)  .
\end{eqnarray}
Here $q_{j}$  are the atomic momenta and $Q=q_{1}+q_{2}+q_{3}$  is
 the total momentum.
The interaction strength $U_{\text{eff}}$  happens to be the same
 function [see Eq.\
(\ref{Ueff})] as in two-body problem. The conventional Faddeev
 reduction
technique (see Ref.\ \cite{Glockle}) leads for three-body bound
 states to
the homogeneous equation
\begin{eqnarray}
X\left( q\right) ={m\over \pi }\int dq^\prime {1\over mE-q^{\prime
 2}-q^\prime q-q^2} \nonumber
\\
\times T_{1D}\left( i\sqrt{m|E|+{3\over 4}q^\prime { } ^{2}}\right)
 X\left( q^\prime \right)  , \label{X}
\end{eqnarray}
where $E<0$ is the three-body bound state energy in the center-of-mass
system and the 1D transition matrix is given by Eq.\ (\ref{T1D}).

\begin{figure}
\includegraphics[width=3.375in]{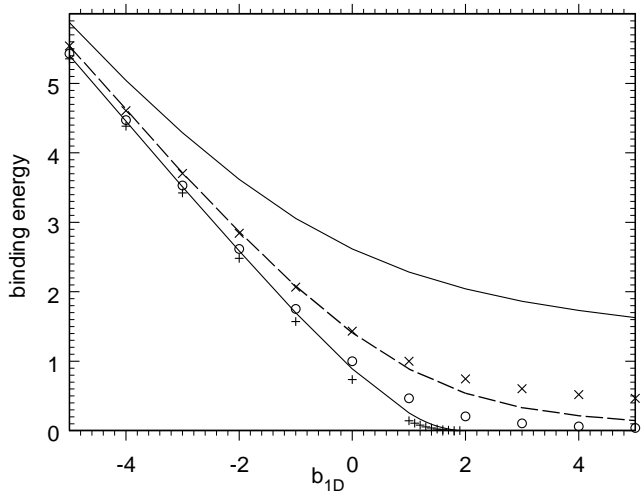}

\caption{The binding energy $E_{b}$  calculated as a function of the
dimensionless detuning (see Eq.\ (\protect\ref{ub1D})). The solid,
dashed, and dot-dashed lines present the energies of three-body bound
states for the dimensionless non-resonant interaction strengths
 $u=1,0$,
and -1, respectively. The corresponding energies of the two-body bound
states are presented by pluses, circles, and crosses, respectively.}
\label{Fig3B}

\end{figure}

Properties of 1D systems depend on two dimensionless parameters:
the non-resonant interaction strength and the detuning, respectively,
\begin{equation}
u=m^{1/3}|g|^{-2/3}U_{a} ,\qquad b_{1D}=m^{-1/3}|g|^{-4/3}D_{1D} .
 \label{ub1D}
\end{equation}
Figure \ref{Fig3B}  presents the scaled binding energy $\epsilon
 _{3}=-
m^{-1/3}|g|^{-4/3}E$ calculated by a numerical solution of Eq.\
(\ref{X}). A related problem with non-resonant interactions
only, the Lieb-Liniger-McGuire model \cite{LL63,McGuire64}, has
an exact solution. The binding energies for two- and three-body
bound states in that model are expressed as $\epsilon _{2}=-u^{2}/2$
 and $\epsilon _{3}=-
2u^{2}$, respectively, and therefore $\epsilon _{3}=4\epsilon _{2}$.
 In the present resonant
case $\epsilon _{3}\approx 4\epsilon _{2}$  only at large positive
 detunings. For large
negative detunings the two-body bound state contains mostly the
1D closed-channel contribution, and the three-body bound state
has a form of the two-body state with a third atom weakly
bounded to it.

\section*{Conclusions}

Two atoms with a Feshbach resonant interaction in an atomic
waveguide can be approximated by a 1D resonant model at low energies
 and
values of the non-resonant scattering length. In the case of a strong
resonance two-atom bound states contain mostly the contributions of
 the
open channel. The closed channel contribution becomes dominant in weak
resonances, such that the 543 G resonance in  $^{6}$Li. In the case of
triatomic molecules a resonant interaction leads to properties
 different
from those predicted by the Lieb-Liniger-McGuire model.

\acknowledgments

The author is very grateful to Maxim Olshanii and Abraham Ben-Reuven
for helpful discussions and to Andrea Simoni for providing parameters
\cite{Simoni05} of the Feshbach resonance in  $^{6}$Li.

\end{document}